\documentclass[letterpaper,12pt]{article}

\usepackage{authblk}
\usepackage{mathptmx}
\usepackage{graphicx}
\usepackage{amsmath}
\usepackage{amssymb}
\usepackage{color}
\usepackage[authoryear,comma,sectionbib]{natbib}

\newcommand{\Prob}{\operatorname{P}}
\newcommand{\given}{\,|\,}
\newcommand{\indep}{\perp\hspace{-0.21cm}\perp}

\newcommand{\argmin}{\operatornamewithlimits{argmin}}

\title{Improving the Efficiency of Genomic Selection}
\author[1]{Marco Scutari}
\author[2]{Ian Mackay}
\author[1]{David Balding}
\affil[1]{{\small Genetics Institute, University College London (UCL), United Kingdom}}
\affil[2]{{\small National Institute of Agricultural Botany (NIAB), Cambridge, United kingdom}}

\begin{document}

\maketitle

\begin{abstract}
We investigate two approaches to increase the efficiency of phenotypic prediction
from genome-wide markers, which is a key step for genomic selection (GS) in plant
and animal breeding. The first approach is feature selection based on Markov
blankets, which provide a theoretically-sound framework for identifying
non-informative markers. Fitting GS models using only the informative markers
results in simpler models, which may allow cost savings from reduced genotyping.
We show that this is accompanied by no loss, and possibly a small gain, in
predictive power for four GS models: partial least squares (PLS), ridge regression,
LASSO and elastic net. The second approach is the choice of kinship coefficients
for genomic best linear unbiased prediction (GBLUP). We compare kinships based
on different combinations of centring and scaling of marker genotypes, and a 
newly proposed kinship measure that adjusts for linkage disequilibrium (LD). 
 
We illustrate the use of both approaches and examine their performances using
three real-world data sets with continuous phenotypic traits from plant and 
animal genetics. We find that elastic net with feature selection and GBLUP 
using LD-adjusted kinships performed similarly well, and were the best-performing
methods in our study.

\vspace{\baselineskip}
\textbf{Keywords:} genome-wide prediction, genomic selection, feature selection,
  Markov blanket, linkage disequilibrium, kinship.
\end{abstract}

\pagebreak

\section{Introduction}

The ever-increasing amount of genetic information available in plant and animal
breeding is reflected in the development of sophisticated models for the prediction 
of quantitative traits from genome-wide markers \citep{heffner,hayes}, also known
as genomic selection (GS). The markers are typically dense single-nucleotide
polymorphisms (SNPs). Approaches to this problem have moved from models with simple 
specifications, such as ridge regression \citep{ridge} and the LASSO \citep{lasso},
to models based on highly-structured hierarchical distributions or semiparametric
approaches. Some examples are the Bayesian alphabet models \citep{bayesalpha,
bayesalpha2}, Bayesian models with complex priors as in \citet{guan-stephens}, 
models based on reproducing kernel Hilbert spaces (RKHS) such as \citet{bravo},
or the Bayesian LASSO \citep{blasso,blasso2}.

This complexity is motivated by the need to correctly model the genetic
architecture of the trait under investigation while producing models that are
easy to estimate even for large SNP profiles. We focus on two key aspects of 
these models: the inclusion of a preliminary step that removes SNPs that appear
to be redundant, and the choice of kinship matrices to model the relatedness of
the genotyped individuals.

The former is equivalent to \emph{feature selection} \citep{sahami}, and can also
be achieved by shrinking SNP effects towards zero, either through the use of
constraints \citep{elastic} or through appropriate prior distributions in a 
Bayesian setting \citep{meuwissen}. We examine the effectiveness in GS of Markov
blankets \citep{pearl}, which have been extensively studied in graphical
modelling. They provide a principled solution to feature selection problems,
and can be implemented as a data pre-processing step prior to fitting the GS
model. We implement Markov blanket feature selection within four GS models
applied to three real-world data sets covering barley, rice and mouse genetics.

Kinship matrices were traditionally derived from pedigrees using a single 
definition, but with kinships now being calculated from SNP data many different
definitions are available \citep{astle}. We investigate four kinship matrices 
within genetic best linear unbiased prediction (GBLUP). These include a novel
matrix introduced by \citet{doug} which adjusts for the bias introduced by 
differences in local linkage disequilibrium (LD), and has been shown to increase
the precision of heritability estimates.

\section{Background}
\label{sec:background}

\subsection{Markov Blankets and Feature Selection}

The Markov Blanket of a variable of interest $T$, denoted as $B(T)$, is the
minimal set of variables conditioned on which all other variables in the model
are probabilistically independent of the target $T$ \citep{pearl}. The Markov
blanket of a phenotype $\mathbf{y}$ in a GS model is the minimal set 
$B(\mathbf{y})\subset \mathbf{X}$ such that
\begin{equation}
\label{def:mb}
  \Prob(\mathbf{y} \given \mathbf{X}) = \Prob(\mathbf{y} \given B(\mathbf{y})), 
\end{equation}
that is, the subset of SNPs $B(\mathbf{y})$ that makes all other SNPs redundant
as far as the trait $\mathbf{y}$ is concerned. Given this property, knowledge of
only the SNPs in $B(\mathbf{y})$ is enough to determine the probability 
distribution of $\mathbf{y}$. Other SNPs become superfluous, either because they
are not associated with the trait or because their effect is mediated by the SNPs
in $B(\mathbf{y})$. If $B(\mathbf{y})$ were known, any GS model could be fitted
using $B(\mathbf{y})$ instead of the full SNP profile $\mathbf{X}$ with no loss
of information, but in practice the need to estimate $B(\mathbf{y})$ means that
some information loss is possible. This two-stage approach contrasts with models
such as BayesB \citep{meuwissen} and the LASSO \citep{lasso}, which select 
significant SNP effects concurrently with model fitting and in a model-specific
way.

Markov blankets can be efficiently estimated from data through the use of conditional
independence tests, such as parametric and non-parametric tests for partial
correlation \citep{legendre,hotelling} or mutual information \citep{pesarin10}.
Tests in common use do not require any tuning parameter except for the type I
error threshold $\alpha$. The estimated $B(\mathbf{y})$ will satisfy (\ref{def:mb})
only approximately because of type I and type II errors. The former arise from
the noisiness inherent to the data and limited sample sizes, while the latter
are typical of weak dependencies which will often be omitted from the Markov
blanket.

Several computationally-efficient heuristic algorithms for Markov blanket
estimation are available in literature, including Grow-Shrink \citep[GS;][]{mphd},
Incremental Association \citep[IAMB;][]{iamb} and Hiton-MB \citep{hitonpc}.
For instance, IAMB can be used to estimate the Markov blanket of a trait 
$\mathbf{y}$ as follows:
\begin{enumerate}
  \item Set $B(\mathbf{y}) = \{ \varnothing \}.$
  \item \textbf{Forward Phase:} until no change is made, 
  \begin{enumerate}
    \item test each SNP $X_i$ for independence from $\mathbf{y}$ conditional on
      the current Markov blanket $B(\mathbf{y})$;
    \item admit into $B(\mathbf{y})$ the SNP whose test returned the lowest
      p-value if that p-value is smaller than $\alpha$.
  \end{enumerate} 
  \item \textbf{Backward Phase:} for each $X_i \in B(\mathbf{y})$, remove $X_i$ 
    from $B(\mathbf{y})$ if $\mathbf{y}$ is independent of $X_i$ conditional 
    on $B(\mathbf{y}) \setminus X_i$.
\end{enumerate}
As a result, conditional independence tests are performed in order of increasing
complexity, thus ensuring that in practice only a small number of SNPs is used
for each test. Compared to single-SNP analyses, such as those described in
\citet{dimauro} and \citet{piepho}, feature selection with Markov blankets is
computationally more expensive because of the use of conditional ($\mathbf{y} 
\indep X_i \given B(\mathbf{y}) \setminus X_i$) instead of marginal ($\mathbf{y} 
\indep X_i$) independence tests. However, as shown in Section \ref{sec:results},
Figure \ref{fig:randomsub}, conditional tests are more effective at discarding
SNPs that carry essentially the same information about the trait and select 
subsets with more predictive power for the same size.

\subsection{Kinship Estimation}
\label{sec:kinship}

In the past, pedigree information was used to specify kinships, but such information
is often missing or inadequate. SNP-based methods for measuring kinships have
become increasingly common and have the advantage of measuring the realised amount
of genome sharing, as opposed to the expected value provided by pedigree-based
methods \citep{astle,forni}

The SNP-based kinship of two individuals is usually based on the average over
SNPs of the product of their genotypes, coded as $0$, $1$ and $2$ according to
the count of one of the two alleles. By design, it can only capture the additive
components of kinship, and it has very low power in identifying non-additive
ones. In the following, we denote this genotype matrix with $\mathbf{X}$, with
rows corresponding to individuals and columns to SNPs, and with $X_i$ its 
$i$th column. 

In human genetics, kinship is commonly measured as the proportion of shared
alleles at each locus \citep{morris}. This approach is also known as 
identical-by-state (IBS) kinship, and will be denoted by $\mathbf{K}_0$. Unlike
other kinship matrices below, $\mathbf{K}_0$ is always non-negative. However,
it cannot be expressed in the form $\mathbf{X} \mathbf{X}^T$, which leads to 
parameters directly interpretable as SNP effect sizes (see Section 
\ref{sec:methods} for details).

Another choice, common in plant and animal genetics, is to centre the genotypes
\citep{habier,vanraden} and estimate the kinship matrix as
\begin{equation}
  \mathbf{K}_1 =  \frac{1}{m}\sum_{i = 1}^m (X_i - 2p_i) (X_i - 2p_i)^T
\end{equation}
where $m$ is the number of markers and $p_i$ is a vector with every entry equal
to the population allele fraction, usually estimated as the mean of $X_i/2$. 
Centring improves interpretability, since kinship values can be interpreted as
an excess or deficiency of allele sharing compared with random allocation of 
alleles, and so zero can be interpreted as ``unrelated''.  However, the 
requirement to estimate the $p_i$, usually from the same data set, can cause
problems in some settings \citep{astle}.

One criticism of both the above choices is that the sharing of a rare allele 
between two individuals counts the same as the sharing of a common allele. 
One natural approach to giving more weight to the sharing of a rare allele is 
to standardise over SNPs, thus obtaining
\begin{align}
\label{eq:allelic}
  &\mathbf{K}_2 = \frac{1}{m}\sum_{i = 1}^m \bar{X}_i \bar{X}_i^T&
  &\text{where}&
  &\bar{X}_i = \frac{X_i - 2p_i}{\sqrt{2p_i(1 - p_i)}}.
\end{align}
The $(i,j)$ entry of $\mathbf{K}_2$ can be interpreted as an average over SNPs
of the correlation coefficient estimated from a single pair of individuals, $i$
and $j$ \citep{astle}.

A modification of $\mathbf{K}_2$ has been recently proposed by \citet{doug},
based on evidence that the effects of SNPs are sensitive to uneven LD across
the genome. In particular, SNP effects are over-estimated in high-LD regions
and under-estimated in low-LD regions due the uneven tagging of causal variants.
The contributions of causal variants are picked up by a larger number of SNPs 
in high-LD regions compared to low-LD regions, thus introducing bias in the GS
models and in turn in subsequent inference such as prediction or heritability
estimation. To correct for this bias, SNPs can be re-weighted:
\begin{equation}
\label{eq:ldak}
  \mathbf{K}_3 = \frac{\sum_{i = 1}^m w_i \bar{X}_i \bar{X}_i^T}{\sum_{i = 1}^m w_i},
\end{equation}
where the weight vector $\mathbf{w} = [w_1 \cdots w_m]$ solves 
\begin{align}
\label{eq:ldadj}
  &\min_{\mathbf{w}} \sum_{i=1}^m | \mathbf{1} - \mathbf{C}_i \mathbf{w}|&
  &\text{subject to}&
  &w_1, \ldots, w_m > 0
\end{align}
and $\mathbf{C}_i$ is a vector of squared correlations of SNP $i$ with
neighbouring SNPs. SNP effects are set to decay exponentially with physical 
distance, according to a decay rate $\lambda$ whose value reflects the average 
LD for the data set. As a result we have that
\begin{equation}
  w_i + \sum_{j \neq i} w_j C_{ij} e^{-\lambda d_{ij}},
\end{equation}
where $d_{ij}$ is the distance between SNPs $i$ and $j$, is approximately 
constant as the weights offset differences in LD as measured by the squared
correlations $C_{ij}$.

For computational reasons, the minimisation in (\ref{eq:ldadj})
is performed separately on different chromosomes and, within each 
chromosome, on different regions chosen based on $\lambda$. 

\section{Materials and Methods}
\label{sec:methods}

We explored the effects of the approaches outlined in Section \ref{sec:background}
on the predictive power of GS models using three publicly-available real-world
data sets including continuous phenotypic traits. The yield data from the AGOUEB
project \citep{agoueb1,agoueb2} consist of $227$ UK winter barley varieties and
$810$ SNPs. The heterogeneous mouse population \citep{mice1,mice2} from the
Wellcome Trust Case Control Consortium (WTCCC) consists of $1940$ SNP profiles
and $12545$ SNPs; among the recorded traits, we consider growth rate and weight.
The rice data set from \citet{rice} consists of $413$ varieties of \emph{Oryza
sativa} with $73808$ SNPs; among the $34$ recorded traits, we consider the number
of seeds per panicle because of its low variability among the various 
subpopulations included in the original analysis.

All data sets have been preprocessed by removing SNPs with minor allele frequencies
$< 1\%$ and those with $> 20\%$ missing data. The missing data in the remaining 
SNPs have been imputed using the \textbf{impute} R package \citep{impute}. Other
widely used imputation methods in genetics, such as that implemented in MaCH 
\citep{mach}, were not available because of the absence of accurate SNP maps;
the position of many SNPs is unknown, and only genetic distances (in cM) were 
available between mapped SNPs. Furthermore, we removed one SNP from each pair
whose allele counts have correlation $> 0.90$ to increase the numerical
stability of the models.

To investigate Markov blanket feature selection, we considered the
following GS models:
\begin{itemize}
  \item Ridge regression, LASSO and the elastic net penalised regressions. 
    These are all based on
    \begin{align}
      &\mathbf{y} = \boldsymbol{\mu} + \mathbf{X}\boldsymbol{\beta} + \boldsymbol{\varepsilon}&
      &\text{with}&
      &\boldsymbol{\hat{\beta}} = \argmin_{\boldsymbol{\beta}} \left\{
         \lambda_1 \|\boldsymbol{\beta}\|_1 + \lambda_2 \|\boldsymbol{\beta}\|_2\right\},
           \lambda_1, \lambda_2 \geqslant 0,
    \end{align}
    where $\mathbf{y}$ is the trait of interest, $\mathbf{X}$ are the SNP
    genotypes, $\boldsymbol{\beta}$ are the fixed SNP effects and 
    $\boldsymbol{\varepsilon}$ are independent, normally-distributed errors
    with variance $\sigma^2_{\varepsilon}$. We used the implementations provided
    by the \textbf{penalized} \citep{penalized} and \textbf{glmnet} \citep{glmnet}
    R packages. When considering the elastic net we restricted both the $L_1$ and
    $L_2$ penalties for the genetic effects $\boldsymbol{\beta}$ to be strictly
    positive ($\lambda_1, \lambda_2 > 0$), to facilitate the comparison with ridge
    regression ($\lambda_1 = 0$) and the LASSO  ($\lambda_2 = 0$).
  \item Partial least squares (PLS) regression as implemented in the \textbf{pls}
    R package \citep{pls}.
  \item Genetic BLUP (GBLUP) implemented in the \textbf{synbreed} R package 
    \citep{synbreed}. It uses the linear mixed model
    \begin{align}
      \label{eq:blup}
      &\mathbf{y} = \boldsymbol{\mu} + \mathbf{Zg} + \boldsymbol{\varepsilon},&
      &\mathbf{g} \sim N(\mathbf{0}, \mathbf{K}\sigma^2_g),
    \end{align}
    where $\mathbf{g}$ are the random effects and $\mathbf{Z}$ is a design
    matrix that can be used for example to indicate the same genotype exposed
    to different environments. Any positive definite matrix can be used for 
    $\mathbf{K}$. Fixed effects can also be included in (\ref{eq:blup}) in
    order to capture purely environmental effects \citep{heffner}.

   When $\mathbf{K}$ can be expressed in the form $\mathbf{XX}^T$, GBLUP can
   be shown to be equivalent to the Bayesian linear regression 
   \begin{align}
     &\mathbf{y} = \sum_{i = 1}^m X^*_i \beta_i + \boldmath{\varepsilon}&
     &\text{with SNP effect prior}&
     &\beta_i \sim N\left(\mathbf{0}, \frac{\sigma^2_g}{m}\mathbf{I}\right),
   \end{align}
   in which $\mathbf{K}$ determines the transformation $X^*$ of the SNP
   genotypes. For instance, the original $X_i$ are used when $\mathbf{K} =
   \mathbf{K}_1$; the scaled $\bar{X}_i$ from (\ref{eq:allelic}) when
   $\mathbf{K} = \mathbf{K}_2$; and the weighted $w_i \bar{X}_i / \sum w_i$
   from (\ref{eq:ldak}) when $\mathbf{K} = \mathbf{K}_3$. This formulation
   of GBLUP results in a more natural interpretation of SNP effects, and is
   sometimes known as random regression BLUP (RR-BLUP). An overview of its
   properties can be found in \citet{rrblup} and \citet{rrblup2}. 

\end{itemize}

Markov blanket feature selection has been performed with the IAMB algorithm as
implemented in the \textbf{bnlearn} R package \citep{jss09}, using the exact
Student's $t$ test for Pearson's correlation with a type I error threshold of 
$\alpha = 0.15$. Each GS model was fitted both using all the available SNPs and
using only the SNPs included in the Markov blanket.

The different kinship matrices were investigated within GBLUP, as the other GS
models do not include an explicit kinship term. $\mathbf{K_1}$ and $\mathbf{K_2}$
were computed using \textbf{synbreed}. For $\mathbf{K_3}$, we used the freely
available LDAK software \citep{ldak}. The LD decay rate was set to $\lambda = 50$cM
for the AGOUEB data, $\lambda = 0.2$cM for the mouse data and $\lambda = 100$cM
for the rice data. Such values were found, through experimentation, to ensure the
LD adjustment was effective without affecting the genetic information present in
the SNP profiles. $\mathbf{K}_0$ was computed with PLINK \citep{plink}. All
configurations of GS models and kinships were fitted once using all SNPs available
after preprocessing the data and once using only those in the Markov blanket.

The predictive power of the GS models was measured with Pearson's correlation
coefficient $\rho$ between the observed trait values and the predictions obtained 
from $10$-fold cross-validation. For each model, cross-validation was run $5$
times. Markov blankets, kinship matrices and GS models were fitted separately
for each fold in each cross-validation run, and the resulting correlations
averaged. The correlation between observed and fitted trait values is also 
reported as a measure of goodness of fit.

\section{Results}
\label{sec:results}

Table \ref{tab:mb} reports the observed correlations ($\rho$, i.e. the correlation
between the observed and the fitted trait values) and the predictive correlations
($\rho_{CV}$, i.e. the correlations obtained from cross-validation) for PLS, ridge
regression, LASSO and the elastic net. The corresponding correlations arising
from the subset of SNPs included in the Markov blankets are labelled $\rho_{MB}$
and $\rho_{CV,MB}$, respectively.

\begin{table}[p]
\small
\begin{center}
\begin{tabular}{l|c|c|c|c|c|c}
  \hline
  Model       & $\rho$ & $\rho_{MB}$ & $\Delta_1$     & $\rho_{CV}$    & $\rho_{CV,MB}$ & $\Delta_2$      \\
  \hline
  \multicolumn{7}{c}{AGOUEB, YIELD ($227$ obs., $185$ SNPs out of $810$, $23\%$) } \\
  \hline
  PLS         & $0.812$          & $0.805$          & $-0.007$ & $0.495$          & $\mathbf{0.495}$ & $+0.000$ \\
  Ridge       & $0.817$          & $0.765$          & $-0.051$ & $\mathbf{0.501}$ & $0.489$          & $-0.012$ \\
  LASSO       & $\mathbf{0.829}$ & $\mathbf{0.811}$ & $-0.018$ & $0.400$          & $0.399$          & $-0.001$ \\
  Elastic Net & $0.806$          & $0.752$          & $-0.054$ & $0.500$          & $0.489$          & $-0.011$ \\
  \hline
  \multicolumn{7}{c}{MICE, GROWTH RATE ($1940$ obs., $543$ SNPs out of $12.5$K, $4\%$) } \\
  \hline
  PLS         & $0.716$          & $0.882$          & $+0.166$ & $0.344$          & $0.388$           & $+0.044$ \\
  Ridge       & $\mathbf{0.841}$ & $0.889$          & $+0.047$ & $0.366$          & $0.394$           & $+0.028$ \\
  LASSO       & $0.717$          & $0.881$          & $+0.164$ & $0.390$          & $0.394$           & $+0.004$ \\
  Elastic Net & $0.751$          & $\mathbf{0.893}$ & $+0.142$ & $\mathbf{0.403}$ & $\mathbf{0.401}$  & $-0.001$ \\
  \hline
  \multicolumn{7}{c}{MICE, WEIGHT ($1940$ obs., $525$ SNPs out of $12.5$K, $4\%$) } \\
  \hline
  PLS         & $\mathbf{0.927}$ & $0.823$          & $-0.104$ & $0.502$          & $0.524$           & $+0.022$ \\
  Ridge       & $0.877$          & $0.843$          & $-0.034$ & $0.526$          & $0.542$           & $+0.016$ \\
  LASSO       & $0.743$          & $0.807$          & $+0.064$ & $0.579$          & $0.577$           & $-0.001$ \\
  Elastic Net & $0.789$          & $\mathbf{0.845}$ & $+0.056$ & $\mathbf{0.580}$ & $\mathbf{0.580}$  & $+0.000$ \\
  \hline
  \multicolumn{7}{c}{RICE, SEEDS PER PANICLE ($413$ obs., $293$ SNPs out of $74$K, $0.4\%$) } \\
  \hline
  PLS         & $0.853$          & $0.923$          & $+0.070$ & $0.583$          & $0.601$           & $+0.018$ \\
  Ridge       & $0.950$          & $0.921$          & $-0.029$ & $0.601$          & $\mathbf{0.612}$  & $+0.011$ \\
  LASSO       & $0.885$          & $\mathbf{0.939}$ & $+0.054$ & $0.516$          & $0.580$           & $+0.064$ \\
  Elastic Net & $\mathbf{0.958}$ & $0.917$          & $+0.040$ & $\mathbf{0.602}$ & $\mathbf{0.612}$  & $+0.010$ \\
  \hline
\end{tabular}

\caption{Correlation coefficients for PLS, ridge regression, LASSO and the
  elastic net: $\rho$ is the correlation between observed and fitted trait
  values; $\rho_{CV}$ is the predictive correlation obtained from cross-validation; 
  $\rho_{MB}$ and $\rho_{CV,MB}$ are the corresponding quantities obtained 
  using only the SNPs in the Markov blanket. $\Delta_1 = \rho_{MB} - \rho$ and 
  $\Delta_2 = \rho_{CV,MB} - \rho_{CV}$. The highest value for each quantity
  and data set is shown in bold. The average dimension of the Markov blanket
  over cross-validation is reported in parentheses for each data set and trait.}
\label{tab:mb}
\end{center}
\end{table}

First of all, we note that for $\alpha = 0.15$ Markov blankets only select a 
small number of SNPs, regardless of the dimension of the SNP profile. The
average size of the Markov blankets obtained from cross-validation is $185$ for
the AGOUEB data, $543$ (for growth rate) and $525$ (for weight) for the mouse
data, and $293$ for the rice data. Of those SNPs, $136$ ($74\%$) appear in at
least half of the cross-validation folds for AGOUEB, $241$ ($46\%$) for the
mouse data and weight, $276$ ($51\%$) for the mouse data and growth rate, but
only $15$ ($5\%$) for the rice data. This can be attributed to the very low
ratio between sample size and number of SNPs in the rice data ($< 0.01$)
compared to the mouse ($0.15$) and AGOUEB ($0.28$) data. As expected, the
dimension reduction is smaller in the case of the AGOUEB data because of the
limited number of available SNPs, despite the extensive LD present in cultivated
UK barley \citep{agoueb2,barleyLD}. On the other hand, only a small proportion
of the original SNPs are retained for the mouse and rice data sets (about $4\%$
and $0.4\%$, respectively). In each case, the number of SNPs included in the
Markov blankets is smaller than the sample size, thus ensuring the regularity
and numerical stability of the GS models.

\begin{figure}[p]
  \includegraphics[width=\textwidth]{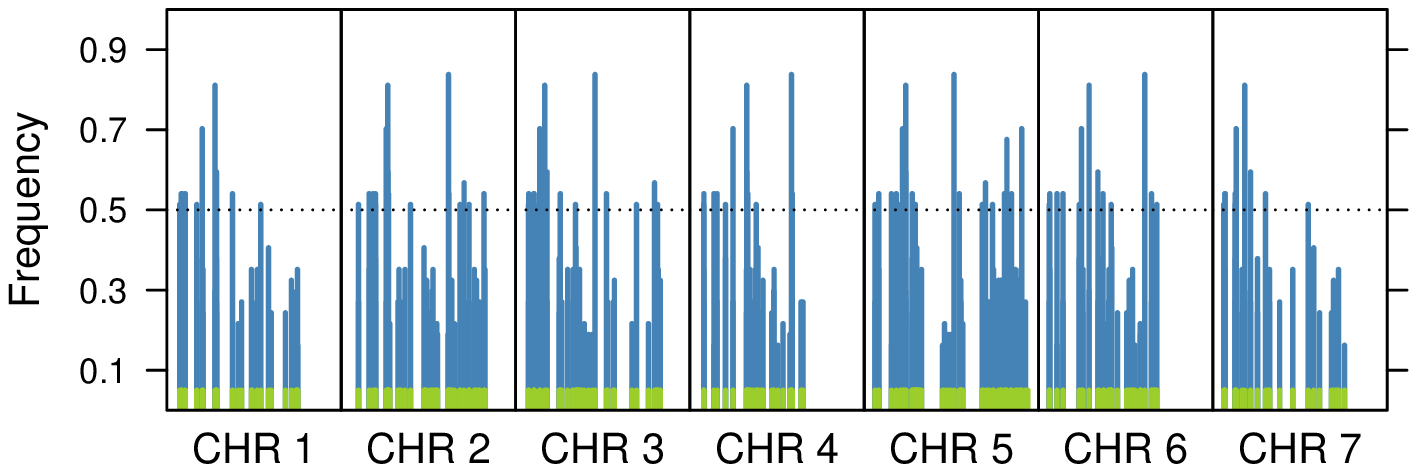}
  \caption{Frequency of the SNPs included in the Markov blankets estimated from
    the AGOUEB data using cross-validation, plotted against the position of the
    SNPs in the barley genome. Green ticks indicate the positions of all mapped
    SNPs for this data set.}
  \label{fig:agouebmap}
  \vspace{\baselineskip}
  \includegraphics[width=\textwidth]{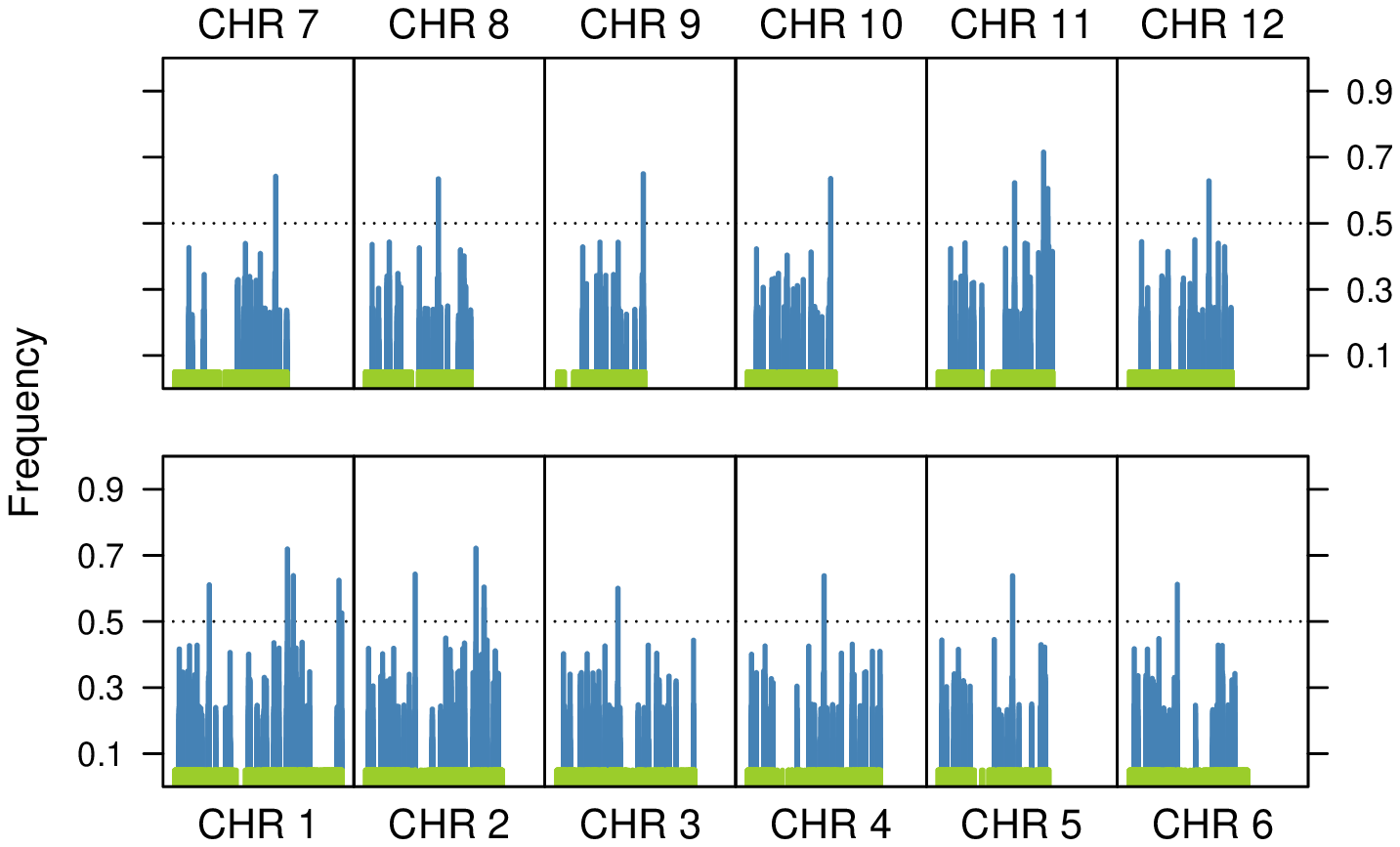}
  \caption{Frequency of the SNPs included in the Markov blankets estimated from
    the rice data using cross-validation, plotted against the position of the
    SNPs in the genome. Green ticks indicate the positions of all mapped SNPs
    for this data set.}
  \label{fig:ricemap}
\end{figure}
\begin{figure}[p]
  \includegraphics[width=\textwidth]{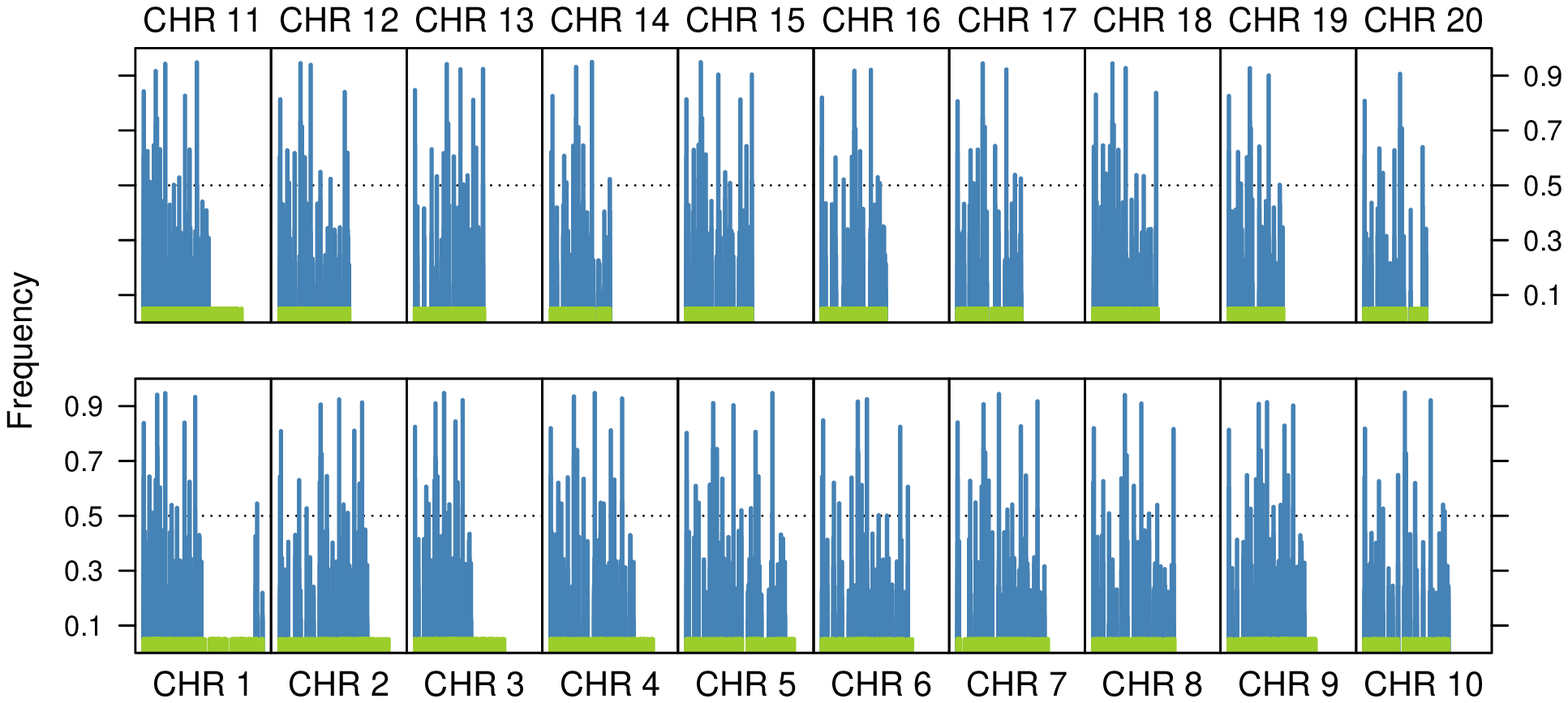}
  \caption{Frequency of the SNPs included in the Markov blankets estimated from
    the mouse weight data using cross-validation, plotted against the position
    of the SNPs in the barley genome. Green ticks indicate the positions of all
    mapped SNPs for this data set.}
  \label{fig:weightmap}
  \vspace{\baselineskip}
  \includegraphics[width=\textwidth]{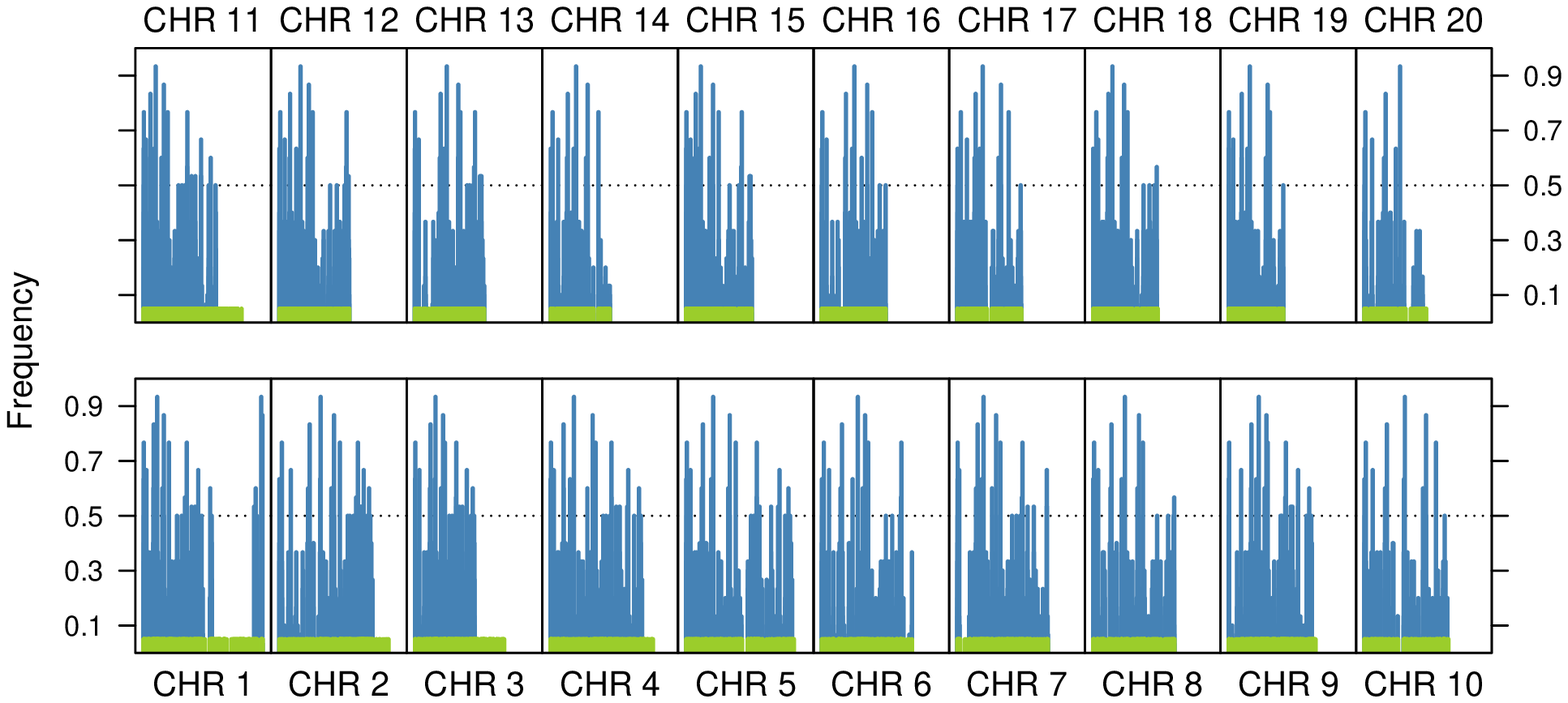}
  \caption{Frequency of the SNPs included in the Markov blankets estimated from
    the mouse growth data using cross-validation, plotted against the position
    of the SNPs in the barley genome. Green ticks indicate the positions of all
    mapped SNPs for this data set.}
  \label{fig:growthmap}
\end{figure}

The position of mapped SNPs within the respective genomes is shown in Figure
\ref{fig:agouebmap} (AGOUEB), Figure \ref{fig:ricemap} (rice), Figure
\ref{fig:weightmap} (mice, weight) and Figure \ref{fig:growthmap} (mice, growth).
For all but the AGOUEB data, we can see how the Markov blankets arising from
cross-validation identify some regions as associated with the trait being
modelled (e.g. SNPs in the range $[0.1\mathrm{cM}, 75.8\mathrm{cM}]$ of 
chromosome $1$ are included with high probability for both traits in the mice
data set) while completely discarding other regions (e.g. $[96.3\mathrm{cM},
108.3\mathrm{cM}]$ in chromosome $2$ and $[70.6\mathrm{cM}, 88.5\mathrm{cM}]$
in chromosome $3$). The positions of these regions may provide useful prior
information in subsequent association studies and in targeting future genotyping
efforts. In the case of the AGOUEB data, marker density is not high enough to
identify regions with markedly different association levels.

\begin{table}[t]
\begin{center}
\small
\begin{tabular}{c|cc|cc|cc|cc}
  \hline
                  & \multicolumn{2}{|c|}{AGOUEB}        & \multicolumn{2}{|c|}{MICE,}         & \multicolumn{2}{|c|}{MICE,}         & \multicolumn{2}{|c}{RICE}           \\
  Kinship         &                  &                  & \multicolumn{2}{|c|}{GROWTH}        & \multicolumn{2}{|c|}{WEIGHT}        &                  &                  \\
   matrix         & $\rho$           & $\rho_{CV}$      & $\rho$           & $\rho_{CV}$      & $\rho$           & $\rho_{CV}$      & $\rho$           & $\rho_{CV}$      \\
  \hline
  $\mathbf{K}_0$  & $\mathbf{0.848}$ & $0.511$          & $\mathbf{0.838}$ & $0.376$          & $\mathbf{0.931}$ & $0.536$          & $\mathbf{0.933}$ & $0.596$          \\
  $\mathbf{K}_1$  & $0.847$          & $0.512$          & $0.656$          & $0.366$          & $0.882$          & $0.507$          & $\mathbf{0.933}$ & $0.590$          \\
  $\mathbf{K}_2$  & $\mathbf{0.848}$ & $0.513$          & $0.688$          & $0.388$          & $0.883$          & $0.508$          & $\mathbf{0.933}$ & $\mathbf{0.598}$ \\
  $\mathbf{K}_3$  & $0.832$          & $\mathbf{0.521}$ & $0.695$          & $\mathbf{0.400}$ & $0.881$          & $\mathbf{0.554}$ & $0.918$          & $0.594$          \\
  \hline
\end{tabular}
\caption{Correlation coefficients obtained in GBLUP using the four kinship
  matrices defined in Section \ref{sec:kinship}. The highest value for each
  quantity and data set is shown in bold. $\rho$ and $\rho_{CV}$ are defined
  as in Table \ref{tab:mb}.}
\label{tab:kinship}
\end{center}
\end{table}

We observe no loss in the predictive power of the GS models after the Markov
blanket feature selection. In fact, the increased numerical stability resulting
from the reduced number of SNPs slightly improved the predictive power of the
GS models. The average of $\rho_{CV}$ over the four analyses was $0.481$, $0.498$,
$0.471$ and $0.521$ for PLS, ridge, LASSO and elastic net respectively, while
the corresponding averages for $\rho_{CV,MB}$ were $0.502$, $0.509$, $0.487$
and $0.520$, all with an approximate standard deviation of $0.0057$ computed
as in \citet{hooper}.

If we choose $\alpha < 0.15$, we obtain Markov blankets that are too small to
capture polygenic effects (results not shown). A possible explanation for this
behaviour may be that large values of $\alpha$ allow Markov blankets to initially
include SNPs that are weakly associated with the trait, to the point that they 
would be individually discarded. In addition, among them there may be sets of
SNPs that are jointly significant due to epistasis, and such sets are retained
in the Markov blanket. 

\begin{figure}[p]
  \includegraphics[width=\textwidth]{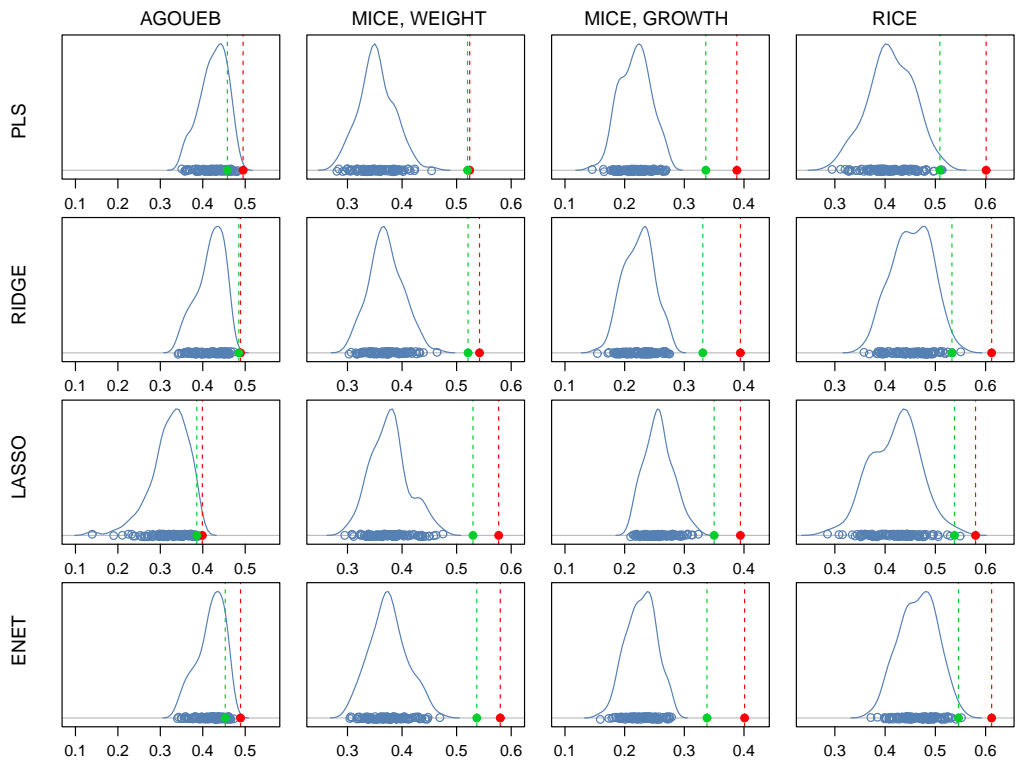}
  \caption{Comparison between the cross-validated correlations obtained from the
    Markov blankets ($\rho_{CV,MB}$, vertical red dashed line in each panel) and
    the subsets of the same size obtained from a single-SNP analysis (green dashed 
    line) and from random sampling (blue empirical density curve).}
  \label{fig:randomsub}
\end{figure}

Furthermore, Markov blankets outperform other subsamples of the same size. To
show this, we generated for each data set $100$ random subsets of SNPs of the
same size as the corresponding Markov blanket. In addition, we also generated
subsamples including the most significant SNPs from a single-SNP analysis under
cross-validation. The same $t$ test as in Markov blanket estimation was used to
assess significance. Subsequently, we used them to fit the GS models 
and to compute the predictive correlations corresponding $\rho_{CV,MB}$. As 
we can see from Figure \ref{fig:randomsub}, the Markov blanket always results 
in higher values of $\rho_{CV,MB}$.

The elastic net consistently outperforms the other GS models both with and 
without the use of Markov blankets, except for the AGOUEB data set (in which
$\rho_{CV}$ is essentially the same for ridge regression and the elastic net).

Overall, from Table \ref{tab:kinship} we see that the predictive performance of
GBLUP improves as the kinship matrices progress from $\mathbf{K}_1$ through to
$\mathbf{K}_3$. $\mathbf{K}_0$, while not being competitive with $\mathbf{K}_3$,
outperforms at least one of $\mathbf{K}_1$ and $\mathbf{K}_2$ for all data sets
but AGOUEB. The means of the four $\rho_{CV}$ values are $0.504$ for $\mathbf{K}_0$,
$0.493$ for $\mathbf{K}_1$, $0.501$ for $\mathbf{K}_2$, and $0.518$ for 
$\mathbf{K}_3$, all with an approximate standard deviation of $0.0057$. Thus,
GBLUP with $\mathbf{K}_3$ performs as well as the elastic net and outperforms
PLS, ridge regression and the LASSO.

\begin{table}[t]
\begin{center}
\footnotesize
\begin{tabular}{c|c|c|c|c|c|c|c|c}
  \hline
  Kinship        & $\rho_{CV,MB}$ & $\Delta_1$ & $\rho_{MB,KIN}$ & $\Delta_2$ & $\rho_{CV,MB}$ & $\Delta_1$ & $\rho_{MB,KIN}$ & $\Delta_2$\\\cline{2-9}
  matrix         &  \multicolumn{4}{c|}{AGOUEB, YIELD}        & \multicolumn{4}{c}{RICE, SEEDS/PANICLE} \\
  \hline
  $\mathbf{K}_0$ & $\mathbf{0.418}$ & $-0.093$ & $0.479$          & $-0.032$ & $0.426$          & $-0.170$ & $\mathbf{0.597}$ & $+0.001$ \\
  $\mathbf{K}_1$ & $0.412$          & $-0.100$ & $0.482$          & $-0.030$ & $0.428$          & $-0.161$ & $0.592$          & $+0.002$ \\
  $\mathbf{K}_2$ & $0.414$          & $-0.099$ & $\mathbf{0.491}$ & $-0.022$ & $\mathbf{0.429}$ & $-0.168$ & $0.589$          & $-0.008$ \\
  $\mathbf{K}_3$ & $0.415$          & $-0.105$ & $0.475$          & $-0.045$ & $0.425$          & $-0.169$ & $0.592$          & $-0.003$ \\
  \hline
                 &  \multicolumn{4}{c|}{MICE, GROWTH RATE}    &\multicolumn{4}{c}{MICE, WEIGHT} \\
  \hline
  $\mathbf{K}_0$ & $0.194$          & $-0.182$ & $0.378$          & $+0.002$ & $0.219$          & $-0.317$ & $\mathbf{0.534}$ & $-0.002$ \\
  $\mathbf{K}_1$ & $0.118$          & $-0.248$ & $0.357$          & $-0.008$ & $0.120$          & $-0.387$ & $0.457$          & $-0.005$ \\
  $\mathbf{K}_2$ & $0.176$          & $-0.211$ & $0.363$          & $-0.025$ & $0.182$          & $-0.326$ & $0.480$          & $-0.028$ \\
  $\mathbf{K}_3$ & $\mathbf{0.195}$ & $-0.204$ & $\mathbf{0.379}$ & $-0.021$ & $\mathbf{0.225}$ & $-0.328$ & $0.530$          & $-0.024$ \\
  \hline
    \end{tabular}
\caption{Correlation coefficients for GBLUP using the four kinship matrices
  defined in Section \ref{sec:kinship} and Markov blanket feature selection.
  $\rho_{CV,MB}$ is defined as in Table \ref{tab:mb}; 
  $\rho_{MB,KIN}$ is the predictive correlation obtained from cross-validation
  with the use of Markov blankets but with the kinship matrices estimated from
  the full SNP profile. The highest value for each quantity and data set is
  shown in bold. $\Delta_1 = \rho_{CV} - \rho_{CV,MB}$ and $\Delta_2 = \rho_{CV} 
  - \rho_{MB,KIN}$, using the $\rho_{CV}$ reported in Table \ref{tab:kinship}.}
\label{tab:both}
\end{center}
\end{table}

Although the elastic net performed equally well with or without Markov blanket
feature selection, that is not the case for GBLUP (Table \ref{tab:both}). For
all kinship matrices, the reduced size of the Markov blanket relative to the full
SNP set detracts from the computation of kinship coefficients, leading to a 
substantial loss of predictive power. If all SNPs are available and can be used
to compute the kinship matrices, then much but not all of this loss is restored.

\section{Conclusions}

We have shown that Markov blanket feature selection applied as a preliminary step
in GS with a continuous trait is able to greatly reduce the size of the SNP set
with no loss (and possibly a small gain) in the predictive power of PLS, ridge
regression, LASSO and the elastic net. Among those models, the elastic net was
the best performer, followed by ridge regression. If GS is to be performed 
repeatedly for the same phenotype, for example in successive generations of crops,
Markov blanket feature selection opens the possibility of reducing costs by
genotyping many fewer markers.

In the absence of a feature selection step, the LD-adjusted kinship matrix 
$\mathbf{K}_3$ \citep{doug} provides slightly better predictive power than the 
matrix with no LD adjustment $\mathbf{K}_2$ \citep{astle} and the IBS kinship
matrix $\mathbf{K}_0$ produced by PLINK \citep{plink}. In turn, $\mathbf{K}_2$
and $\mathbf{K}_0$ appear superior to the matrix with neither LD adjustment
nor standardising of SNPs $\mathbf{K}_1$ \citep{habier}. Using $\mathbf{K}_3$,
GBLUP was competitive with the elastic net (both had mean $\rho_{CV} = 0.52$
over the four datasets).

Markov blanket feature selection is not compatible with GBLUP because of the
requirement for large numbers of SNPs to compute the kinship matrix. However,
Markov blanket feature selection has only a small adverse effect on GBLUP
if all SNPs are available for computing the kinship matrix.

\section*{Acknowledgements}

The work presented in this paper forms part of the MIDRIB project, which is funded
by the UK Technology Strategy Board (TSB) and Biotechnology \& Biological Sciences 
Research Council (BBSRC),  grant TS/I002170/1.  We thank our project partners for
helpful discussions. We also thank the AGOUEB Consortium (supported by UK DEFRA,
the Scottish Government, through the Sustainable Arable LINK Program Grant 
302/BB/D522003/1) for making their data available.

\end{document}